\documentclass[aps,prl,superscriptaddress,twocolumn]{revtex4}
\usepackage{graphics}
\usepackage{epsfig}
\usepackage{color}
\usepackage{stackrel}
\usepackage{amsfonts}
\usepackage{wrapfig}
\usepackage{pstricks}
\usepackage{pst-node}
\usepackage{bm}
\usepackage{dcolumn}
\usepackage{multirow}
\usepackage{epstopdf}
\usepackage{subfigure}
\usepackage{amssymb}
\usepackage{amsmath}
\usepackage{commath}
\usepackage{graphicx,bm}
\usepackage{verbatim}
\usepackage{booktabs}
\usepackage[para]{threeparttable}
\usepackage{etoolbox}
\usepackage{lipsum}

\begin{document}
\title{Exciton valley depolarization in monolayer transition-metal dichalcogenides}

\author{Min~Yang}
\affiliation{Department of Electrical and Computer Engineering, University of Rochester, Rochester, New York 14627, USA}

\author{Cedric Robert}
\affiliation{Universit\'{e} de Toulouse, INSA-CNRS-UPS, LPCNO, 135 Av. Rangueil, 31077 Toulouse, France}

\author{Zhengguang Lu}
\affiliation{National High Magnetic Field Laboratory, Tallahassee, FL 32310, USA}
\affiliation{Physics Department, Florida State University, Tallahassee, FL 32306, USA}
\author{Dinh~Van~Tuan}
\affiliation{Department of Electrical and Computer Engineering, University of Rochester, Rochester, New York 14627, USA}
\author{Dmitry Smirnov}
\affiliation{National High Magnetic Field Laboratory, Tallahassee, FL 32310, USA}
\author{Xavier Marie}
\affiliation{Universit\'{e} de Toulouse, INSA-CNRS-UPS, LPCNO, 135 Av. Rangueil, 31077 Toulouse, France}
\author{Hanan~Dery}
\altaffiliation{hanan.dery@rochester.edu}
\affiliation{Department of Electrical and Computer Engineering, University of Rochester, Rochester, New York 14627, USA}
\affiliation{Department of Physics and Astronomy, University of Rochester, Rochester, New York 14627, USA}

\begin{abstract}
The valley degree of freedom is a sought-after quantum number in monolayer transition-metal dichalcogenides. Similar to optical spin orientation in semiconductors, the helicity of absorbed photons can be relayed to the valley (pseudospin) quantum number of photoexcited electrons and holes. Also similar to the quantum-mechanical spin, the valley quantum number is not a conserved quantity. Valley depolarization of excitons in monolayer transition-metal dichalcogenides due to long-range electron-hole exchange typically takes a few ps at low temperatures. Exceptions to this behavior are monolayers MoSe$_2$ and MoTe$_2$ wherein the depolarization is much faster. We elucidate the enigmatic anomaly of these materials, finding that it originates from Rashba-induced coupling of the dark and bright exciton branches next to their degeneracy point. When photoexcited excitons scatter during their energy relaxation between states next to the degeneracy region, they reach the light cone after losing the initial helicity. The valley depolarization is not as fast in monolayers WSe$_2$,  WS$_2$ and likely MoS$_2$ wherein the Rashba-induced coupling is negligible.
 \end{abstract}

\maketitle

For half a century, optical orientation has been a ubiquitous approach to study the spins of electrons and holes in semiconductors \cite{Lampel_PRL68,OO_85,Kikkawa_PRL98,Dzhioev_PRB02,Hilton_PRL02,Zutic_RMP04,Dyakonov_Book,Li_PRL10,Pezzoli_PRB13}. The spin-orbit interaction allows one to use the angular momentum of absorbed photons to orient the spins of photoexcited electron-hole pairs  \cite{OO_85,Zutic_RMP04}. One can then probe the ensuing spin relaxation through the circular-polarization decay of the emitted photons. This approach received intense attention with the discovery of monolayer transition-metal dichalcogenides (ML-TMDs) \cite{Splendiani_NanoLett10,Mak_PRL10,Korn_APL11,Zeng_NatNano12,Mak_NatNano12,Feng_NatComm12,Jones_NatNano13,Xu_NatPhys14,Mak_NatPhot16,Wang_RMP18}, in which time-reversal symmetry and the lack of space inversion symmetry lock the valley and spin degrees of freedom \cite{Xiao_PRL12,Song_PRL13}. 

Upon excitation of ML-TMDs with a circularly polarized light, the photon angular momentum is transferred to the helicity of the exciton. The helicity carries information on the identity of the valley in which the optical transition took place \cite{Xiao_PRL12}. Similar to the spins of electrons and holes, the valley degree of freedom is not a conserved quantity and excitons lose their original helicity over time \cite{Wang_RMP18}. Experiments show that the valley polarization of optically-active (bright) excitons typically decays within few ps \cite{Lagarde_PRL14,Zhu_PRB14,Wang_PRB14,DelConte_PRB15,Yan_SR15,Plechinger_NatCom16,Schmidt_NanoLett16,Huang_PRB17}, and theory shows that the decay is induced by the electron-hole exchange interaction \cite{Yu_NatComm14,Glazov_PRB14,Yu_PRB14,Yu_NSR15,Glazov_PSSB15,Baranowski_2DMater17}. The exceptions are ML-MoSe$_2$ and ML-MoTe$_2$ for which photoluminescence  experiments show negligible  circular
polarization degree indicating a much faster spin/valley depolarization \cite{MacNeill_PRL15,Wang_APL15,Robert_PRB16,Kioseoglou_SR16,Tornatzky_PRL18}. To date, the physical origin of this anomaly remained a conundrum. 

The focus of this Letter is on analyzing the exciton valley depolarization in ML-TMDs and understanding the reason for the anomaly of ML-MoSe$_2$ and ML-MoTe$_2$. In addition to the long-range electron-hole exchange that couples bright excitons with opposite helicity, we consider the spin-orbit-coupling between bright and spin-forbidden (dark) excitons \cite{Dery_PRB15}. When an exciton traverses through the two-dimensional crystal, it experiences a fluctuating Rashba potential induced by local out-of-plane electric fields due to ripples, strained regions, defects inside the ML, or charged impurities in the surrounding dielectric layers \cite{supp}.  These fields strongly mix the bright and dark exciton states if they are nearly degenerate. 

\begin{figure}
\includegraphics[width=7.5cm,height=5cm]{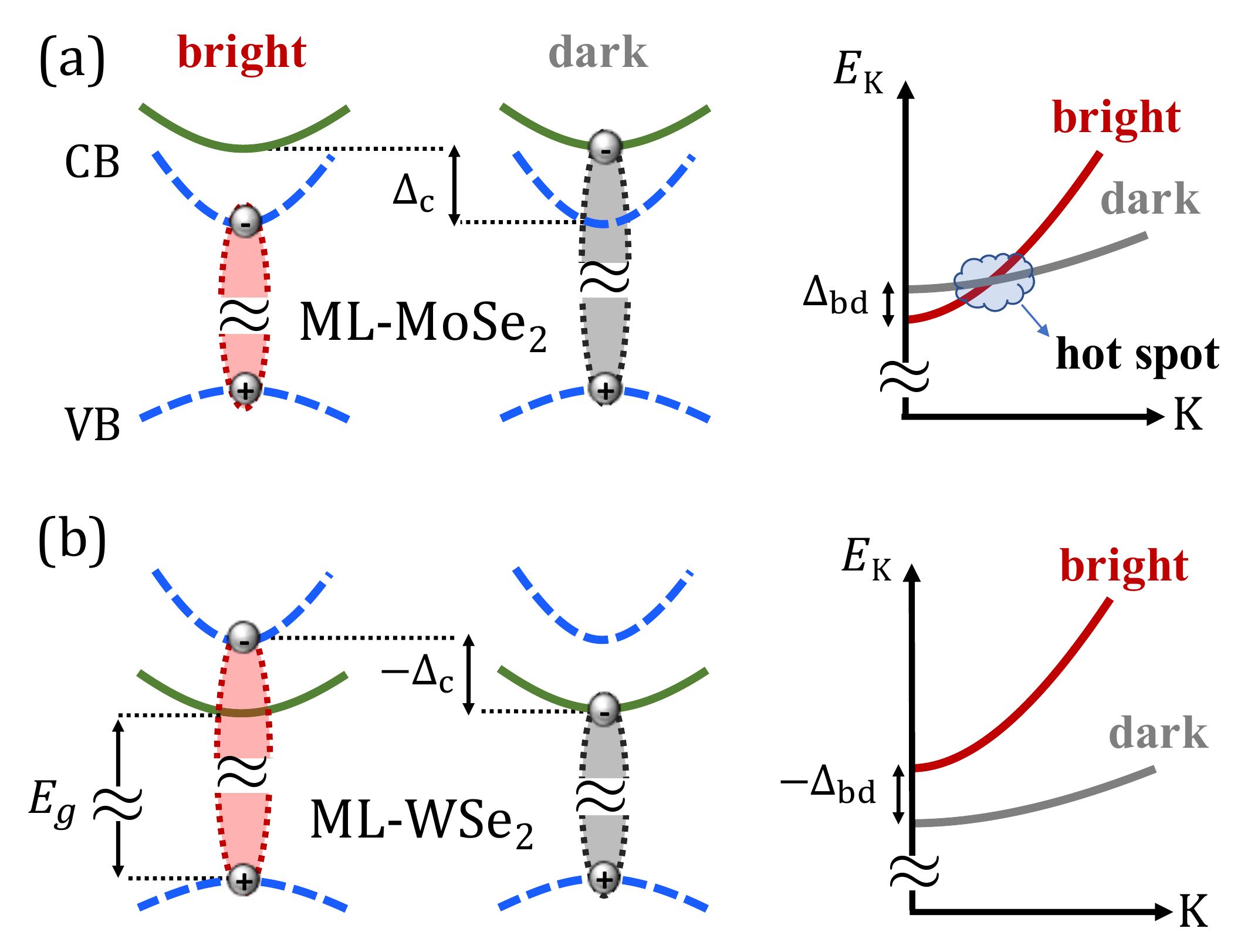}
 \caption{Schematics of excitons in (a) ML-MoSe$_2$ and (b) ML-WSe$_2$. Bright excitons are formed when the spin of the missing electron in the valence band (VB) matches that of the electron in the conduction band (CB). In both materials and other ML-TMDs, the electron component of the dark exciton is from the CB valley with heavier mass. The right schemes show the energy-wavevector dispersion relations of excitons in the absence of electron-hole exchange and Rashba interactions. $\Delta_{\text{bd}}$ is the bright-dark exciton energy splitting, governed by their different binding energies and the spin-splitting energy of the CB, $\Delta_{\text{c}}$.  
 } \label{fig:scheme}
\end{figure}

We identify a few important phenomena by using Monte Carlo simulations to quantify the exciton energy relaxation process and calculate the ensuing polarization evolution of excitons that spontaneously radiate from the light cone. First, the valley depolarization mostly takes place before photoexcited excitons reach thermal equilibrium with the lattice. Second, the depolarization is strongly enhanced due to the Rashba-type mixing of bright and dark exciton states next to their degeneracy point. This phenomenon can be viewed as  a hot spot in the exciton dispersion, as shown in Fig.~\ref{fig:scheme}(a), and it is applicable in ML-MoSe$_2$ or ML-MoTe$_2$. Finally, the exciton state mixing is strongest and the ensuing depolarization time is fastest when the bright and dark excitons are nearly degenerate at the light cone, $\Delta_{\text{bd}} \rightarrow 0$ in Fig.~\ref{fig:scheme}, or when the energy of the dark exciton is just a few meV above that of the bright one. For example, using the recently measured energy difference between the dark and bright excitons of ML-MoSe$_2$, $\Delta_{\text{bd}}=+1.5$~meV \cite{Lu_arXiv19}, we calculate a nearly complete exciton valley depolarization in less than 1 ps.  In comparison and in agreement with experimental results, we find slower depolarization in ML-WSe$_2$  where the energy difference is $\Delta_{\text{bd}}=-40$~meV \cite{Zhang_NatNano17,Zhou_NatNano17,Wang_PRL17}.

In what follows, we first analyze the energy relaxation of photoexcited excitons in ML-TMDs and then present a model that explains the Rashba-induced coupling between bright and dark excitons during the energy relaxation process. Lastly, we discuss the results and benchmark the findings against experimental results. 

\begin{figure}
\includegraphics[width=6cm]{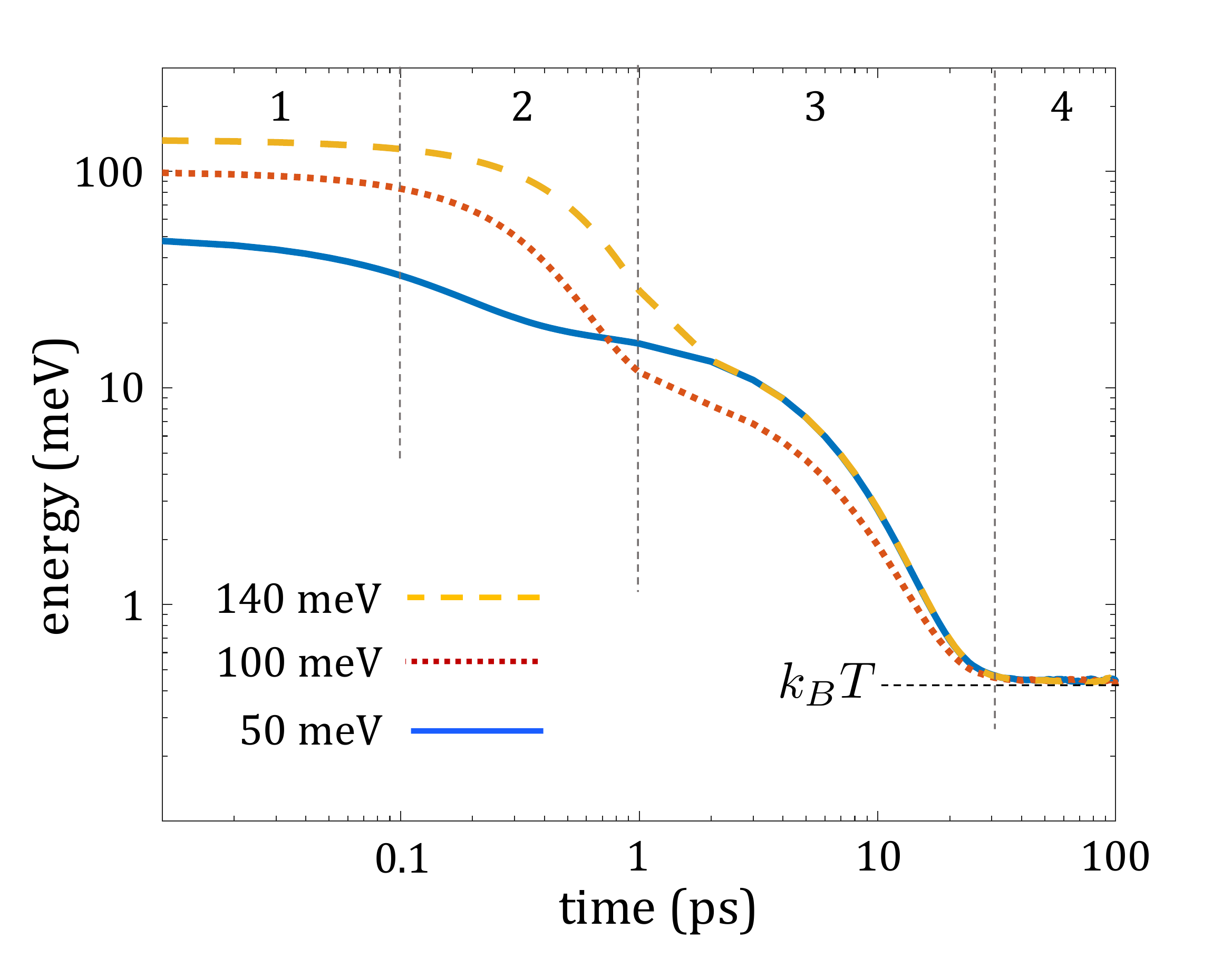}
 \caption{Monte Carlo simulation of the exciton energy relaxation in ML-MoSe$_2$ at T=5~K. Results are shown for three initial kinetic energies. The excitons reach thermal equilibrium with the lattice after nearly 30 ps. The relaxation includes four steps, separated by vertical dashed lines (see text).} \label{fig:energy_relaxation}
\end{figure}

The energy relaxation process is studied through Monte Carlo simulations of 10$^5$ hot excitons. The relaxation is governed by interaction of the excitons with long-wavelength phonons. The mechanisms included are the long-range Fr\"{o}hlich interaction with longitudinal-optical (LO) phonons, the short-range interaction with homopolar phonons (can be viewed as electron and/or hole interactions with thickness fluctuations), and the deformation-potential interaction with acoustic phonons \cite{Kaasbjerg_PRB12,Sohier_PRB16,Thilagam_JAP16,Shree_PRB18,VanTuan_PRL19,VanTuan_arXiv19}.  The Supplemental Material includes technical details of these simulations. Figure~\ref{fig:energy_relaxation} shows the energy relaxation evolution of bright excitons in ML-MoSe$_2$ at 5~K for three different kinetic energies of the initial hot exciton population. These results do not change qualitatively in other ML-TMDs \cite{supp}. Note that the electron-hole exchange and Rashba interaction have not been introduced yet (i.e., the exciton branches are not coupled). 

Figure~\ref{fig:energy_relaxation} shows four steps in the low-temperature energy relaxation of hot excitons that are introduced at $t=0$ with three different initial kinetic energies. The first step is the coherent regime before the first scattering and it lasts during the first 0.1~ps. The second step  is dominated by emission of homopolar phonons, and it typically ends $\sim$1~ps after photoexcitation. Unlike the energy relaxation of electrons or holes, the Fr\"{o}hlich interaction with neutral excitons is relatively weak due to the similar mass of electrons and holes: the strong interaction of the electron with the macroscopic polarization induced by the LO phonon is offset by the respective interaction of the hole \cite{supp,VanTuan_PRL19}. The third step takes place when the average exciton energy is below that of the optical phonon ($\sim$30 meV). The excitons are still hot and their relaxation is governed by emission of acoustic phonons. The duration of this process is 20-30~ps,  in agreement with recent measurements in high quality ML-MoSe$_2$ \cite{Fang_PRL19}. Finally, the excitons reach thermal equilibrium with the lattice wherein the exciton-phonon interaction has similar probabilities to emit and absorb phonons. By this time, a few tens ps after photoexcitation, time-resolved experiments reveal that the valley polarization has already  decayed \cite{Lagarde_PRL14,Zhu_PRB14,Wang_PRB14,DelConte_PRB15}. Thus, calculating the valley depolarization by assuming thermal exciton distribution oversimplifies the experimental conditions. 

Next we introduce the Hamiltonian of the exciton system, and later we will introduce a model that combines its eigenstates with the Monte Carlo simulation results. The Hamiltonian of bright and dark excitons reads \cite{Dery_PRB15}
\begin{equation}
\mathcal{H}(\mathbf{k}) =  \left( \begin{array}{cc} H_b & H_R \\ H_R^{\ast} & H_d  \end{array} \right). \label{eq:H}
\end{equation}
 $\mathbf{k} = k(\cos{\theta}, \sin{\theta})$ is the two-dimensional center-of-mass wavevector (crystal momentum) of the exciton. The upper diagonal block belongs to bright excitons \cite{MacNeill_PRL15,Yu_NatComm14},
\begin{equation}
H_b =  \frac{\hbar^2k^2}{2M_b}\mathcal{I} + J_0k(\cos{2\theta} \sigma_x + \sin{2\theta}\sigma_y) \,. \label{eq:Hb}
\end{equation}
$M_b$ is the bright-exciton mass and $J_0$ is the long-range electron-hole exchange parameter \cite{Maialle_PRB93}. $\mathcal{I}$ is the 2$\times$2 identity matrix and $\sigma_i$ are the Pauli matrices. The lower diagonal block in Eq.~(\ref{eq:H}) is of dark excitons,
\begin{equation}
H_d =   \Delta_{bd} + \frac{\hbar^2k^2}{2M_d}\mathcal{I}  + J_d \sigma_z\,.\label{eq:Hd}
\end{equation}
$M_d$ is the dark-exciton mass and $\Delta_{bd}$ is the bright-dark energy splitting at the light cone. The last term includes the short-range exchange interaction of dark excitons \cite{Dery_PRB15,Slobodeniuk_2DMater16},  which is not relevant for our discussion on accounts of its small value, $J_d \lesssim 1$~meV \cite{Robert_PRB17,Molas_arXiv19}. Finally, the off-diagonal block in Eq.~(\ref{eq:H}) is the Rashba coupling between bright and dark excitons \cite{Dery_PRB15},
\begin{equation}
H_R =  \alpha_R k E_z \left( \begin{array}{cc} \exp(-i\theta)  & \exp(-i\theta)  \\  - \exp(i\theta)  & \exp(i\theta)   \end{array} \right). \label{eq:HR}
\end{equation}
$E_z$ is the out-of-plane electric field and $\alpha_R$ is the Rashba coefficient. In general, the value of $\alpha_R$ for excitons is larger than that of thermal electrons \cite{Kormanyos_PRX14}. The small exciton size means that its wavefunction is spread in momentum space. Therefore, the exciton wavefunction includes electron and hole states with wavevector components away from the valley center, wherein the effect of remote bands on the spin mixing is evident \cite{Dery_PRB15}. 

To evaluate the valley depolarization, we first denote the eigenstates of $\mathcal{H}(\mathbf{k})$ in Eq.~(\ref{eq:H}) by $| n, \mathbf{k} \rangle$ where $n$ is the index of one of the four possible states. The trivial basis states $| \ell \rangle$, where $| \ell=1 \rangle = [1,0,0,0]^T$ , ... , $| \ell=4 \rangle = [0,0,0,1]^T$, are then expressed by the superposition 
\begin{equation}
| \ell \rangle =  \sum_n C_{\ell,n} (\mathbf{k}) | n, \mathbf{k} \rangle\, ,  \label{eq:ell_nk}
\end{equation}
where  $C_{\ell,n} (\mathbf{k}) \equiv \langle  n, \mathbf{k}   | \ell \rangle$.  The probability that the exciton superposition state evolves from $| \ell \rangle$ at time $t$ to $| j \rangle$ at time $t+\tau$, reads
\begin{equation}
\left| \langle \ell_t | j_{t+\tau} \rangle \right|^2 =   \left| \sum_n C_{\ell,n}^{\ast}(\mathbf{k}) C_{j,n}(\mathbf{k}) e^{-iE_n(\mathbf{k})\tau/\hbar} \right|^2 .\label{eq:Pij}
\end{equation}
$E_n(\mathbf{k})$ are the eigenvalues of $\mathcal{H}(\mathbf{k})$. Assuming the initial helicity of the exciton is $\sigma^+$ (i.e., $| \ell=1 \rangle$ at $t=0$), we update the 4-component polarization vector after each scattering according to the probabilities in Eq.~(\ref{eq:Pij}) \cite{supp}. To do so, we employ the values of $\tau$ and $\mathbf{k}$  from the Monte Carlo simulations, where $\tau$ is the time between exciton-phonon scattering events and $\mathbf{k}$ is the exciton wavevector during this time. The circular polarization degree, $p(t)$, is then given by the difference between the first and second components of the polarization vector at time $t$. This physical picture is similar to the evolution of the electron spin polarization during Dyakonov-Perel relaxation in noncentrosymmetric semiconductors \cite{Dyakonov_SPSS72}. 

To model the measured effect seen in experiments, the results we present below are achieved by collecting $4\times 10^5$  (rare) events in which excitons manage to get to the light cone and radiate spontaneously. We have checked that collecting more radiative events do not change the results and conclusions. The initial kinetic energy of a photoexcited exciton is randomized by following a Gaussian distribution, $\mathcal{N}(E_0,\sigma^2)$, where $E_0$ is the mean and $\sigma = 5$~meV is the standard deviation. The latter simulates the broadening due to pulse excitation conditions or energy uncertainty caused by the geminating exciton-phonon process when the photoexcitation is outside the light cone. Excitons then relax in energy by emitting phonons, where the vast majority end their life non-radiatively before reaching the minuscule light cone (see Supplemental Material for details). The exchange parameter we employ in the simulation is $J_0=10$~meV$\cdot$\AA, which is similar to the value suggested in \cite{Glazov_PRB14}, and it leads to very good agreement with the decay times observed in time-dependent experiments \cite{Lagarde_PRL14,Zhu_PRB14,Wang_PRB14,DelConte_PRB15}. 

\begin{figure}
\includegraphics[width=8.5cm]{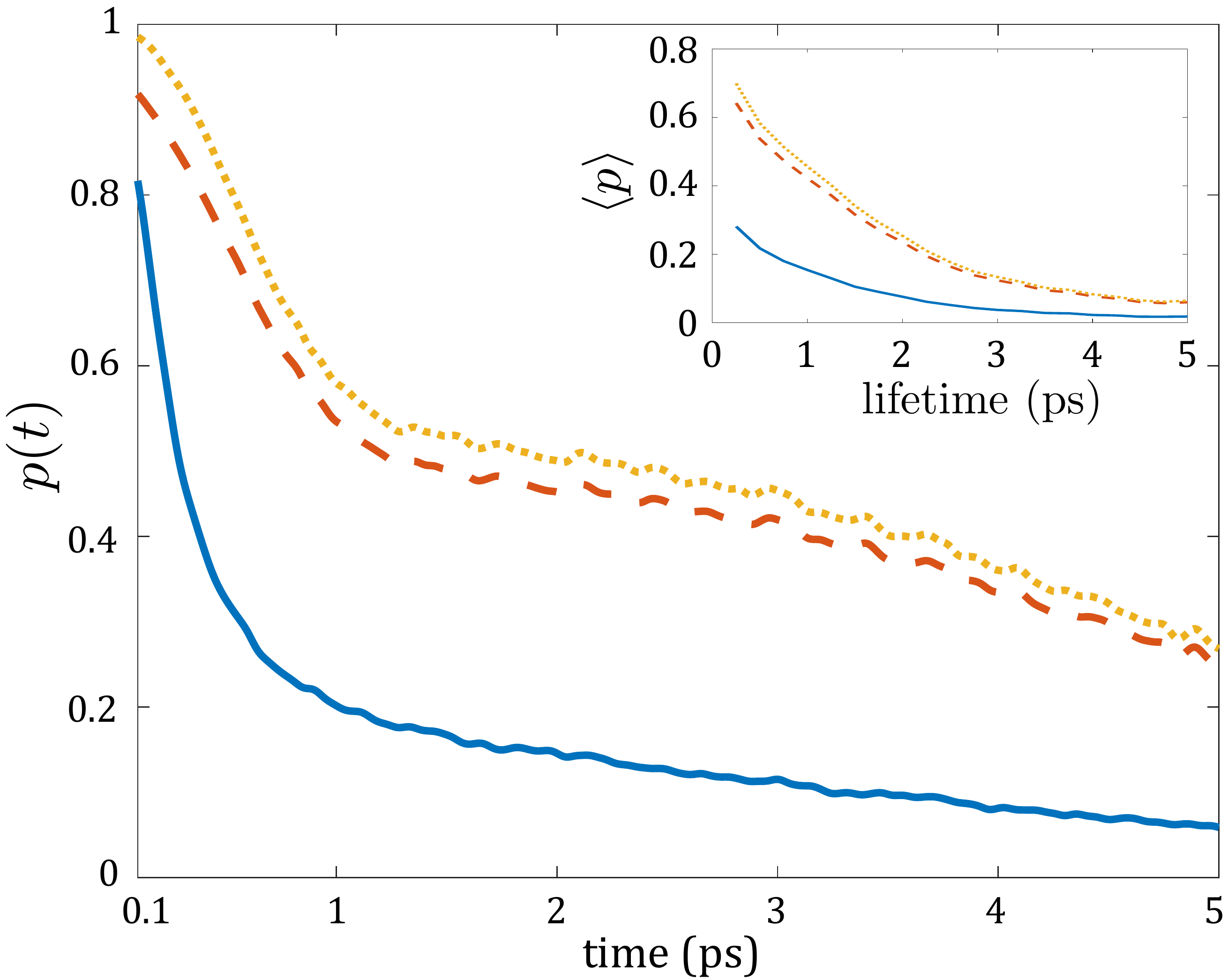}
 \caption{The exciton valley depolarization at T=5~K when $E_0=100$~meV and $J_0=10$~meV$\cdot$\AA. The dotted line shows the depolarization evolution of bright excitons without the Rashba coupling to the dark excitons. Depolarization with the Rashba coupling when $\Delta_{\text{bd}}=-40$~meV ($+$1.5~meV) is shown by the dashed (solid) line. Inset: Average polarization as a  function of the exciton lifetime.} \label{fig:theory}
 \end{figure}

Figure~\ref{fig:theory} shows the calculated valley depolarization of radiative excitons for three cases. The first calculation, denoted by the dotted line, is without the Rashba interaction [i.e., $E_z= 0$ in Eq.~(\ref{eq:HR})]. The second and third ones are with the Rashba interaction, calculated with $\Delta_{\text{bd}}=-40$ (dashed line) and $+$1.5~meV (solid line). The  amplitude of the Rashba parameter was randomized uniformly in the range $0 < \alpha_R E_z < 0.05$~eV$\cdot$\AA$\,$ with each scattering. This choice imitates the fluctuating out-of-plane electrical fields that excitons experience over time when they traverse the crystal. Clearly, Fig.~\ref{fig:theory} shows that the  Rashba interaction is relevant when the bright and dark exciton branches are nearly degenerate; i.e., when $\Delta_{\text{bd}}$ is small. The faster depolarization in this limit is reminiscent of the ultrafast spin relaxation of holes in unstrained bulk semiconductors wherein each scattering between the degenerate or nearly degenerate heavy and light hole states leads to significant spin relaxation due to the spin-mixed hole states \cite{Hilton_PRL02,Pezzoli_PRL12}.

Next, we calculate the average polarization as a function of the exciton lifetime. Non-radiative recombination processes or transitions to lower-energy states, such as trions, control the hot excitons lifetime and limit the time during which excitons should reach the light cone and recombine radiatively. Accordingly, a shorter exciton lifetime leads to larger average polarization at the expanse of smaller quantum yield (less radiative events). The inset of Fig.~\ref{fig:theory} shows the calculated average polarization, $ \langle p \rangle = \int_0^\infty dt p(t) f(t)$, where $f(t)$ is the probability density function and $p(t)$ is the exciton valley depolarization function [main part of Fig.~\ref{fig:theory}]. The Monte Carlo simulations were used to generate $f(t)$ from the photoluminescence intensity profile \cite{supp}. The parameters and mechanisms that govern $f(t)$ and $p(t)$ are the initial exciton energy and exciton scattering rates. In addition, $p(t)$ is strongly influenced by the electron-hole exchange and Rashba interaction, whereas $f(t)$  by the exciton lifetime. 

\begin{figure}
\includegraphics[width=8.5cm]{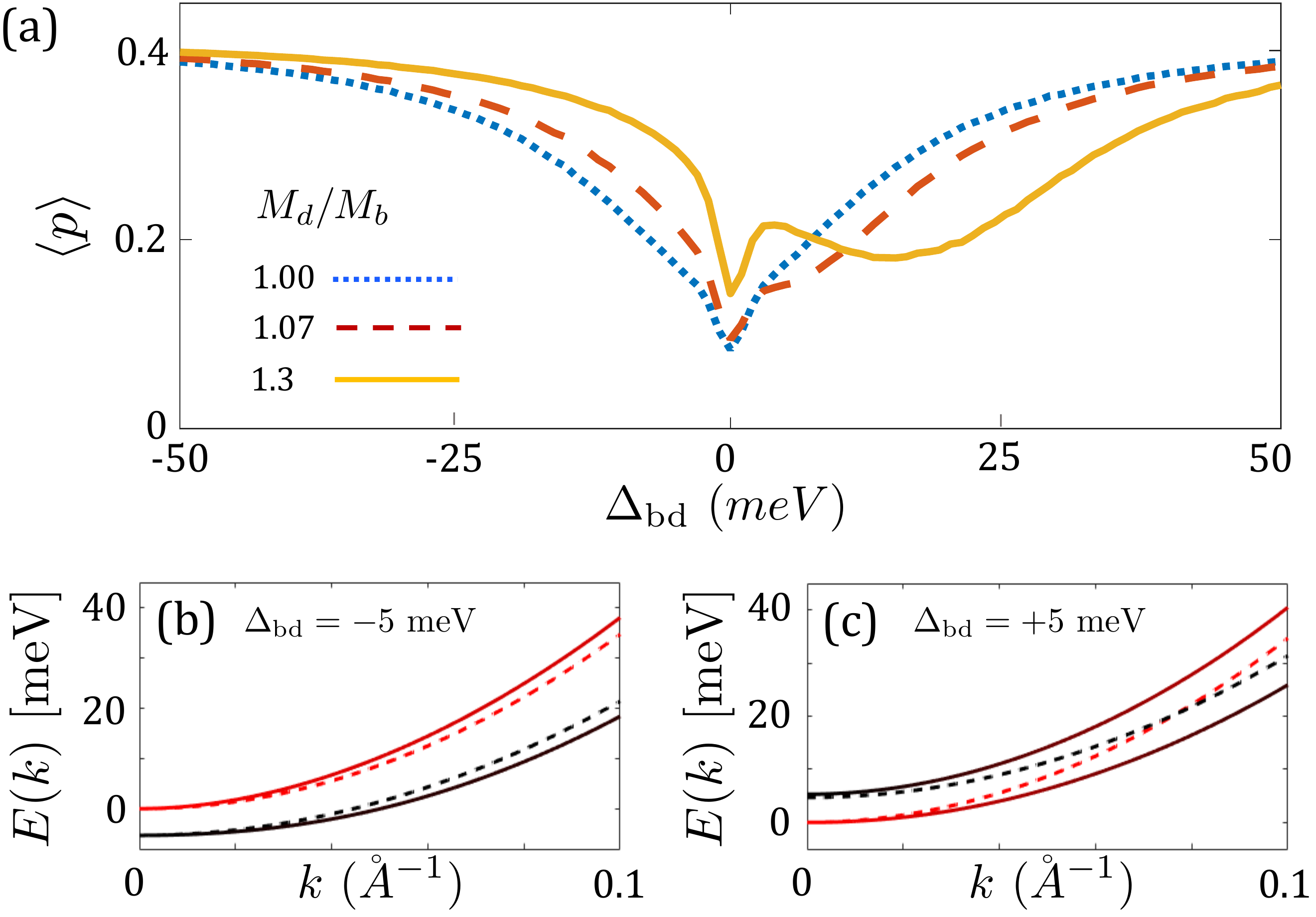}
 \caption{ (a) The average exciton polarization as a function of $\Delta_{\text{bd}}$ when the initial exciton kinetic energy is 100~meV, the exciton lifetime is 1.5~ps and the temperature is 5~K. The three lines show the results of different ratios between the dark and bright exciton masses.  The signature of the hot spot becomes evident when the mass ratio increases and $\Delta_{\text{bd}}>0$. The main change to these results by choosing a longer/shorter exciton lifetime is smaller/larger polarization values on the $y$-axis. (b) and (c) show the dark and bright exciton dispersion when $\Delta_{\text{bd}}=\mp 5$~meV and $M_d/M_b = 1.3$. The solid and dashed lines are the results for $\alpha_R E_z = 50$ and 0~meV$\cdot$\AA, respectively.} \label{fig:dbd}
\end{figure}

Finally, we calculate the average polarization while treating $\Delta_{\text{bd}}$ as a free parameter. Figure~\ref{fig:dbd}(a) shows results when the exciton lifetime is 1.5~ps for three ratios between the masses of the dark and bright excitons. We can identify two distinct features. The first one is the polarization dip when $\Delta_{\text{bd}} \rightarrow 0$. The second feature is the asymmetry in the polarization between positive and negative values of $\Delta_{\text{bd}}$, where this effect is pronounced when dark excitons become increasingly more heavy than the bright ones. The polarization increases rapidly away from zero  when $\Delta_{\text{bd}}$ is negative compared with the case that it is positive. 

The calculated behavior in Fig.~\ref{fig:dbd}(a) can be understood by inspecting the energy dispersion relations of bright and dark excitons. Figures~\ref{fig:dbd}(b) and (c) compare the cases when $\Delta_{\text{bd}}=\pm5$~meV for $M_d/M_b=1.3$. The dashed lines are the dispersion relations without the Rashba interaction, where we can see that the two branches cross (depart)  when $\Delta_{\text{bd}}$ is positive (negative). The red and black colors denote bright and dark excitons, respectively. When the energy dispersion is calculated in the presence of the Rashba interaction, as shown by the  solid lines, a clear avoided crossing behavior emerges when $\Delta_{\text{bd}}>0$. Figures~\ref{fig:dbd}(c) corresponds to the case that $\Delta_{\text{bd}} = +5$~meV, where bright excitons belong to the lower branch in the light cone ($k \rightarrow 0$)  and to the upper branch when the exciton kinetic energy is larger than a few tens meV. This behavior can be traced by the redness of the solid lines in (c), calculated from the weight of the bright components in the eigenstate. The avoided crossing is far less evident in (b) where $\Delta_{\text{bd}} = -5$~meV. In this case, the bright and dark excitons remain in the same branch. Viewing the Rashba interaction as a fluctuating field, an exciton entering a region with a relatively large field can experience either diabatic or adiabatic passage between the two branches. The diabatic transition is relevant when $\Delta_{\text{bd}}$ is small and positive (strong avoided crossing) whereas the adiabatic transition is relevant in other cases. 

The results of Figs.~\ref{fig:theory} and \ref{fig:dbd} demonstrate the importance of $\Delta_{\text{bd}}$. This parameter has contributions from three sources, $\Delta_{\text{bd}}=\Delta_{c}+\Delta_x+\Delta_m$. The first contribution is from the spin-orbit interaction in the conduction band, $\Delta_{c}$, giving rise to a relatively small energy splitting between the top and bottom valleys (up to few tens meV) \cite{Kosmider_PRB13,Cheiwchanchamnangij_PRB13,Kormanyos_2DMater15}.  Here, we assume that the monolayers are undoped and therefore neglect the effect on $\Delta_{c}$ due to many-body exchange interactions \cite{Dery_PRB16,VanTuan_PRX17,VanTuan_PRB19,Scharf_JPCM19}. The second contribution to $\Delta_{\text{bd}}$ is due to the repulsive short-range electron-hole exchange interaction, $\Delta_{x}<0$,  which raises the energy of the bright exciton compared with the dark one \cite{Echeverry_PRB16,Deilmann_PRB17,Zhang_NatNano17}. The third contribution stems from the mass difference between the bright and dark excitons, $\Delta_m$. This contribution is also negative because of the increased binding energy of the dark exciton in ML-TMDs: its electron component comes from the  conduction-band valley with heavier effective mass \cite{Kormanyos_2DMater15}.

To the best of our knowledge, there are no conclusive measurements of $\Delta_{\text{bd}}$ in ML-MoTe$_2$ and ML-MoS$_2$. In light of this fact, we turn to ab-initio calculations which show that the change in $\Delta_{\text{bd}}$ between the molybdenum-based monolayers is dominated by the change in $\Delta_{c}$ \cite{Echeverry_PRB16}. The latter is governed by the competition between the transition-metal and chalcogen atoms, where increasing the mass of the chalcogen (transition-metal) atom `pushes' the value of $\Delta_{c}$ to be positive (negative) \cite{Echeverry_PRB16,Deilmann_PRB17,Kormanyos_2DMater15}.  Using $\Delta_{\text{bd}}=+1.5$~meV in ML-MoSe$_2$ as a reference point \cite{Lu_arXiv19}, we therefore assume that $\Delta_{\text{bd}}$ is somewhat larger in ML-MoTe$_2$ whereas it is likely to become negative in ML-MoS$_2$. Combining this assumption with the results of Fig.~\ref{fig:dbd}, we find consistency with the observations that the valley depolarization in ML-MoTe$_2$ is much faster than in ML-MoS$_2$, where the latter resembles the cases of ML-WSe$_2$ and ML-WS$_2$ wherein $\Delta_{\text{bd}}$ is negative \cite{MacNeill_PRL15,Wang_APL15,Robert_PRB16,Kioseoglou_SR16,Tornatzky_PRL18}.

In conclusion, we have identified important depolarization mechanisms that can elucidate the origin of the minute circular polarization degree observed in photoluminescence experiments of ML-MoTe$_2$ and ML-MoSe$_2$. Whereas the depolarization of bright excitons due to the long-range electron-hole exchange interaction is expected to be similar in all ML-TMDs, the Rashba-type coupling between bright and dark exciton provides an additional valley depolarization process under certain conditions. We have shown that when the Rashba interaction leads to pronounced avoided crossing between the branches of bright and dark excitons, the valley depolarization is much enhanced. Consistent with the empirical findings that the valley depolarization is weaker in ML-WSe$_2$, ML-WS$_2$ and ML-MoS$_2$, we find that the avoided crossing is a relatively weak effect in these compounds. By improving the understanding of the exciton dynamics in ML-TMDs, we hope that implications of this work will lead to better control of the sough-after valley degree of freedom in these compounds.

\acknowledgments{We thank Mikhail Glazov for fruitful discussions. This work was mainly supported by the Department of Energy, Basic Energy Sciences, under Contract No. DE-SC0014349. The computational work (Monte Carlo simulation) was also supported by the National Science Foundation (Grant No. DMR-1503601). Cedric Robert and Xavier Marie acknowledge funding from ANR 2D-vdW-Spin, ANR VallEx and ANR MagicValley. Xavier Marie also acknowledges the Institut Universitaire de France. The work performed at the National High Magnetic Field Laboratory was supported by the National Science Foundation Cooperative Agreement No. DMR-1644779 and the State of Florida.}

\begin{widetext}

\section{Supplemental information: Exciton valley depolarization in monolayer transition-metal dichalcogenides}

The information in this document  includes:

\begin{enumerate}

\item Exciton-phonon matrix elements and a compiled list of the parameter values we use in the simulations

\item Monte Carlo simulations and calculation of valley depolarization

\end{enumerate}


\section{E\lowercase{xciton-phonon interactions in} ML-TMD\lowercase{s}} 

Of the nine phonon modes at the zone center of ML-TMDs, six belong to the optical branches. Two of which are strongly coupled to spin-conserving scattering of electrons or holes \cite{Kaasbjerg_PRB12, Song_PRL13}: The longitudinal  optical (LO) and out-of-plane transverse optical (ZO) phonons. The LO mode is denoted by  $E_2'$ (or $\Gamma_6$) and the ZO by $A_1'$ (or $\Gamma_1$). Here we consider the Fr\"{o}hlich interaction with the LO mode and the short-range scattering due to thickness fluctuations induced by the ZO mode. Both are considered in the long-wavelength limit. In addition, we consider the interaction between exciton and acoustic phonons due to an effective deformation potential that lumps together the contributions from longitudinal- and transverse-acoustic modes (LA and TA). Below, we describe the electron (or hole) interaction with these phonon modes. 

The matrix element of exciton interaction with optical phonons contains the interaction terms $D_{j,\lambda}(\mathbf{q})$ where $\mathbf{q}$ is the phonon wavevector, $j=e(h)$ represents the electron (hole) component of the exciton, and $\lambda =\{ E_2', A_1'\}$. The Fr\"{o}hlich interaction due to coupling with the mode $E_2'$  is governed by the coupling parameter  \cite{Sohier_PRB16}
\begin{equation}
D_{E_2'}(q)= D_{e,E_2'}(q)= D_{h,E_2'}(q) =  \sqrt{n_{E_2'} + \frac{1}{2} \pm \frac{1}{2} } \sqrt{\frac{\hbar^2A_u}{2AM_xE_{E_2'}}}  \left( 1 + \sqrt{\frac{M_x}{M_m} }\right) \frac{2\pi Z_{E_2'} e^2}{A_u\epsilon_{3\chi}(q)}   \,\,\,, \label{eq:LO}
\end{equation}
where $n_{E_2'} = 1/[\text{exp}(E_{E_2'}/k_BT)-1]$ is the Bose-Einstein distribution. $E_{E_2'}$ is the phonon energy where we have neglected its weak dependence on $\mathbf{q}$ due to the dispersionless nature of long-wavelength optical phonons. The $\pm$ denotes the case of phonon emission (plus) or absorption (minus). $A$ and $A_u$ are the areas of the ML and unit cell, respectively.  $M_x$ and $M_m$ are the masses of the chalcogen and transition-metal atoms, respectively. $Z_{E_2'}$ is the Born effective charge describing the linear relation between the force on the atom and the macroscopic electric field. Conservation of charge implies that $Z_{E_2'}=Z_{m}=2Z_x$. $\epsilon_{3\chi}(q)$ is the static dielectric function and can be found in Ref.~\cite{VanTuan_PRB18}.  

The coupling of electrons and holes to ZO phonons ($A_1'$ mode) is governed by the short-range potential induced by the volume change of the unit-cell volume. This coupling can be viewed as the scattering that electrons or holes experience due to thickness fluctuations of the ML in the long-wavelength limit.  The corresponding interaction terms read \cite{Song_PRL13,Sohier_PRB16}
\begin{equation}
D_{j,A_1'}(q) \simeq D_{j,A_1'} = \sqrt{n_{A_1'} + \frac{1}{2} \pm \frac{1}{2} } \sqrt{ \frac{\hbar^2 A_u}{2A(2M_x+M_m) E_{A_1'} }}  \mathcal{S}^{(A_1')}_{j} \,\,\,, \label{eq:ZO}
\end{equation}
where the Bose-Einstein distribution in this case is, $n_{A_1'} = 1/[\text{exp}(E_{A_1'}/k_BT)-1]$, and as before, we have neglected the weak $\mathbf{q}$-dependence of the phonon energy ($E_{A_1'}$) due to the dispersionless nature of long-wavelength optical phonons. $\mathcal{S}^{(A_1')}_{j}$ is the scattering constant of electrons ($j=e$) or holes ($j=h$).

The coupling of excitons to acoustic phonons $D_{j,\text{ac}}(q)$ is governed by the deformation potential $\Xi_j$ \cite{Kaasbjerg_PRB12,Song_PRL13}
\begin{equation}
D_{j,\text{ac}}(q) = \sqrt{n_{\text{ac}} + \frac{1}{2} \pm \frac{1}{2} } \sqrt{ \frac{\hbar^2 A_u}{2A(2M_x+M_m) \hbar v_s q }}  \Xi_j q \,\,\,. \label{eq:ac}
\end{equation}
where $n_{\text{ac}} = 1/[\text{exp}(\hbar v_s q/k_BT)-1]$ and $v_s$ is the effective sound velocity. 

Assuming weak coupling between excitons and phonons, the corresponding matrix element  reads
\begin{eqnarray} 
M_{\lambda}(\mathbf{K}_2,\mathbf{K}_1; \mathbf{q}) = \left\langle   \Psi_X(\mathbf{r_h}, \mathbf{r_e}; \mathbf{K}_2)   |  D_{e,\lambda}(q) e^{i\mathbf{q}\mathbf{r}_{e}}  - D_{h,\lambda}(q) e^{i\mathbf{q}\mathbf{r}_{h}}  |    \Psi_X(\mathbf{r_h}, \mathbf{r_e}; \mathbf{K}_1)   \right\rangle.  \label{eq:M_general}
\end{eqnarray} 
$\mathbf{K}_{2(1)}$  is the exciton wavevector in the final (initial) state, 
\begin{eqnarray} 
\Psi_X(\mathbf{r_h}, \mathbf{r_e}; \mathbf{K}) &=& \frac{\exp({i\mathbf{K}{\mathbf{R}}})}{\sqrt{A}}\varphi(r) \,\,, \qquad \mathbf{r}=\mathbf{r_e}-\mathbf{r_h} \,\,, \qquad \qquad \qquad  \mathbf{R}=  \beta_e \mathbf{r}_e + \beta_h \mathbf{r}_h \,\, . \label{eq:psi}
\end{eqnarray} 
$\varphi(r)$ is the exciton ground state (1$s$ state), $\mathbf{R}$ and $\mathbf{r}$ are the center-of-mass and relative coordinates, $\beta_e = m_e/(m_e+m_h)$ and $\beta_h = 1-\beta_e$. Substituting Eq.~(\ref{eq:psi}) in (\ref{eq:M_general}), the translation symmetry dictates that ($\mathbf{q}=\mathbf{K}_2-\mathbf{K}_1$)
\begin{eqnarray} 
M_{\lambda}(\mathbf{K}_2,\mathbf{K}_1; \mathbf{q}) \equiv M_{\lambda,\mathbf{q}} = \left\langle   \varphi(r)   |  D_{e,\lambda}(q) e^{i\beta_h \mathbf{q}\mathbf{r}}  - D_{h,\lambda}(q) e^{i\beta_e\mathbf{q}\mathbf{r}}  |    \varphi(r)   \right\rangle.  \label{eq:M_2}
\end{eqnarray} 
We have used the stochastic variational method (SVM) to express $\varphi(r)$ in terms of correlated Gaussians \cite{VanTuan_PRB18,Varga_CPC08,Varga_PRC95,Mitroy_RMP13,Kidd_PRB16,Donck_PRB17,VanTuan_PRL19,VanTuan_arXiv19}, 
\begin{eqnarray} 
\varphi(r) &=& \sum_{j=1}^n C_j \exp\left( -\tfrac{1}{2}\alpha_{j} r^2     \right) \,\,,\label{eq:svm_form}
\end{eqnarray} 
where $n$ is the number of correlated Gaussians needed to accurately describe the ground state. Using this wavefunction form, we can perform the integration over $\mathbf{r}$ analytically. That is, $M_{\lambda,\mathbf{q}}$ becomes a discrete sum over elements that are expressed in terms of the (real) variational parameters, $C_j$ and $\alpha_{j}$. Given that it is sufficient to use a few tens of correlated Gaussians to accurately describe the exciton states, the calculation of  Eq.~(\ref{eq:M_2}) is efficient and fast \cite{VanTuan_arXiv19},
\begin{eqnarray} 
 M_{\lambda,\mathbf{q}} =   \sum_{i,j}^n  \frac{2\pi C_i C_j}{\alpha_i+\alpha_j}   \left[ D_{e,\lambda}(q) \exp\left(-\frac{\beta_h^2q^2}{2(\alpha_i+\alpha_j)}\right)  - D_{h,\lambda}(q) \exp\left(-\frac{\beta_e^2q^2}{2(\alpha_i+\alpha_j)}\right) \right].  \label{eq:M_3}
\end{eqnarray}
Finally, the scattering rate is calculated from the Fermi Golden rule,
\begin{eqnarray} 
\left( \tau_{\lambda, K,\pm} \right)^{-1} = \frac{2\pi}{\hbar}\sum_{\mathbf{q}}|M_{\lambda,\mathbf{q}}|^2  \delta\left(   E_{\mathbf{K}}  - (E_{\mathbf{K}-\mathbf{q}} \pm E_{\lambda,\bm{q}})   \right), \label{eq:FGR}
\end{eqnarray}
where $E_K= \hbar^2 K^2/2(m_e+m_h)$  is the kinetic energy of the exciton prior to scattering. Phonon emission is denoted by the plus sign and absorption by the minus sign. Given that the energies of the optical phonons are of the order of a few tens meV in all ML-TMDs, only the spontaneous phonon emission is relevant for the optical modes at low temperatures (i.e., the Bose-Einstein distributions $n_{A_1'}$ and $n_{E_2'}$ are negligible). 

\subsection{The parameter values we use in the simulations}
The SVM calculation of the exciton ground state and dielectric function $\epsilon_{3\chi}(q)$ for the hBN encapsulated monolayer are exactly the same as in Refs.~\cite{VanTuan_PRB18}, \cite{VanTuan_PRL19} and \cite{VanTuan_arXiv19}. In addition,
\begin{enumerate}
\vspace{-1mm}
\item The area of the unit cell is $A_u = \sqrt{3}a_{lc}^2/2 = 8.87~\AA^2$ where  $a_{lc}=3.2~\AA$~ is the triangular lattice constant. 
\vspace{-1mm}
\item The atomic masses of molybdenum, tungsten and selenium are $M_m=M_{\text{Mo}}=1.59\cdot 10^{-22}$~g, $M_m=M_{\text{W}}=3.05\cdot 10^{-22}$~g and $M_{\text{Se}}=M_x=1.31\cdot 10^{-22}$~g.
\vspace{-1mm}
\item The effective masses of the electron and hole are $m_e=0.5m_0$ and $m_h=0.6m_0$ in ML-MoSe$_2$ and  $m_e=0.29m_0$ and $m_h=0.36m_0$ in ML-WSe$_2$ \cite{Kormanyos_2DMater15}.
\vspace{-1mm}
\item The optical-phonon energies are $E_{E_2'}=35$~meV and $E_{A_1'}=29.8$~meV in ML-MoSe$_2$, $E_{E_2'}=32$~meV and $E_{A_1'}=31$~meV in ML-WSe$_2$ \cite{VanTuan_PRL19}.
\vspace{-1mm}
\item The Born effective charges are $Z_{E_2'}=-1.16$ in ML-WSe$_2$ and $Z_{E_2'}=-1.78$ in ML-MoSe$_2$  \cite{Sohier_PRB16}.
\vspace{-1mm}
\item The sound velocities are $v_{s}=4.1\times 10^5$~cm/s in ML-MoSe$_2$ and $v_{s}=3.3\times 10^5$~cm/s in ML-WSe$_2$ \cite{Jin_PRB14}.
\vspace{-1mm}
\item The scattering constants due to thickness fluctuations are  $\mathcal{S}^{(A_1')}_{e}=10$~eV/\AA ~ and $\mathcal{S}^{(A_1')}_{h}=5$~eV/\AA ~ in both ML-MoSe$_2$ and ML-WSe$_2$. We note that DFT calculations in the literature report different results \cite{Sohier_PRB16,Kaasbjerg_PRB12,Jin_PRB14}, where all show that the scattering constants are of the order of a few eV per \AA. We have used a large difference between the electron and hole scattering constants in order to affect the relaxation of excitons. Such a strong difference was evident in the DFT calculations of Ref.~[\onlinecite{Sohier_PRB16}]. 
\vspace{-1mm}
\item The deformation potentials are $\Xi_e^{\text{ac}}=7.6$~eV and $\Xi_h^{\text{ac}}=1.8$~eV in ML-MoSe$_2$, and $\Xi_e^{\text{ac}}=6.5$~eV and $\Xi_h^{\text{ac}}=1.1$~eV in ML-WSe$_2$. These values follow the analysis of Shree \textit{et al.} who fit the deformation potential parameters to match the line-shape of the PL due to the interaction of excitons with long-wavelength acoustic phonons \cite{Shree_PRB18}. These values are larger than the ones calculated by  DFT \cite{Kaasbjerg_PRB12}. Note that the use of larger values offsets the fact that elastic scattering of excitons off impurities has been neglected  (which becomes a relevant scattering after the excitons thermalize and slow down).
\end{enumerate}

Figure~\ref{fig:tau}(a) and (b) show the scattering rates in ML-WSe$_2$ and  ML-MoSe$_2$ at T=5~K as a function of the exciton kinetic energy. The Fr\"{o}hlich coupling provides the weakest relaxation channel in spite of the fact that its coupling constant is relatively large (i.e.,  $D_{j,E_2'}(q) > D_{j,A_1'}(q)$). The reason is that the electron and hole component cancel each other effectively when $D_{e,E_2'}= D_{h,E_2'}$, whereas $D_{e,A_1'} \neq D_{h,A_1'}$. As a result, while the transport of electrons or holes is dominated by the Fr\"{o}hlich interaction at elevated temperatures, the energy relaxation of hot excitons in ML-TMDs is dominated by short-range scattering with thickness fluctuations. The energy relaxation of excitons due to the Fr\"{o}hlich interaction can only become relevant if $m_e \gg m_h$ (or $m_e \ll m_h$) so that $\beta_e \rightarrow 1$ and $\beta_h \rightarrow  0$ (or vice versa) in Eq.~(\ref{eq:M_3}). 

Figure~\ref{fig:tau}(a) and (b) show that the scattering time of hot excitons due to the short-range thickness fluctuations is in the ballpark of a few hundreds fs. When the exciton energy is smaller than the optical phonon energy, the relaxation is governed by emission of acoustic phonons. The scattering time is in the ballpark of a few ps. Phonon emission ceases at very small exciton kinetic energies, and phonon absorption becomes dominant as shown in panel (c).  

\begin{figure*}
\includegraphics[width=17cm]{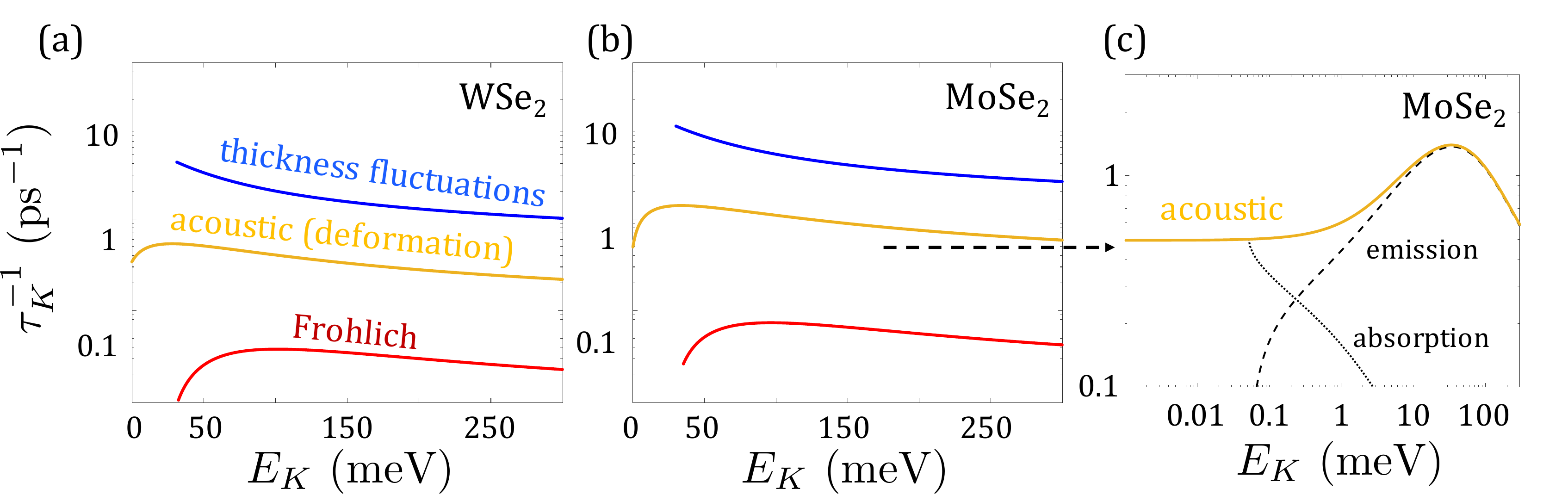}
 \caption{Scattering rates in ML-WSe$_2$ (a) and  ML-MoSe$_2$ (b) at T=5~K as a function of the exciton kinetic energy, $E_K$. (c) Breakup of the scattering rate into phonon emission and absorption components in the case of exciton interaction with acoustic phonons in ML-MoSe$_2$.} \label{fig:tau}
\end{figure*}

\section{M\lowercase{onte}-C\lowercase{arlo Simulations}}

We explain the  time-of-flight concept and then describe how the simulations are performed. We start by defining the overall scattering time of an exciton whose wavevector amplitude is $K$,
\begin{eqnarray} 
\frac{1}{\tau_K} &\equiv&  \frac{1}{\tau_{E_2', K,+}} + \frac{1}{\tau_{A_1', K,+}} + \frac{1}{\tau_{\text{ac}, K,+}} + \frac{1}{\tau_{\text{ac}, K,-}} + \frac{1}{\tau_{\text{r}, K}} + \frac{1}{\tau_{\text{nr}, K}}. \label{eq:tauK_total}
\end{eqnarray} 
The first and second rates on the right-hand side are due to scattering events that involve emission of optical phonons (Fr\"{o}hlich and thickness fluctuations), the third and fourth are  due to scattering events that involve emission and absorption of acoustic phonons, and the fifth and sixth rates are due to radiative and non-radiative recombination. The scattering with phonons were defined in Eq.~(\ref{eq:FGR}) with the help of Eqs.~(\ref{eq:LO})-(\ref{eq:ac}) and (\ref{eq:M_3}). The radiative and non-radiative recombination rates are defined as 
\begin{eqnarray} 
\frac{1}{\tau_{\text{r}, K}} = \frac{H(E_{\text{lc}} - E_K )}{\tau_{\text{r},0}}\,\,, \qquad  \frac{1}{\tau_{\text{nr}, K}} = \frac{\exp(-E_K/E_{\ell})}{\tau_{\text{nr},0}}\,\,.\label{eq:tau_rnr}
\end{eqnarray} 
The intrinsic radiative recombination time is $\tau_{\text{r},0}=0.1$~ps, and it is relevant only when excitons are in the minuscule light cone, as indicated by the Heaviside step function $H( E_{\text{lc}} - E_K)$.  We have used $E_{\text{lc}}=0.01$~meV.  The intrinsic non-radiative recombination time, $\tau_{\text{nr},0}$, is referred to as the exciton lifetime in the main text. We have used $\tau_{\text{nr},0}=1.5$~ps in Figs.~3 and 4 of the main text, whereas it is a variable in the inset of Fig.~3.  The non-radiative process is relevant when excitons are not too energetic, and we have used that $E_K \lesssim E_{\ell}=100$~meV.  The reason for choosing these time constants is that they reproduce the time-resolved PL intensity profile seen in experiments \cite{Robert_PRB16}. More about what happens when we change these time constants is explained in Fig.~\ref{fig:pANDf} and Sec.~\ref{sec:figs}.

\vspace{2mm}


Next, we define the maximal scattering rate
\begin{eqnarray} 
\frac{1}{\tau_{\text{m}}} = \text{max}\left\{     \frac{1}{\tau_K}   \right\} \,\,. \label{eq:taum}
\end{eqnarray}
Among all $K$ values of $\tau_K$ in Eq.~(\ref{eq:tauK_total}), $\tau_{\text{m}}$ is the fastest scattering time.

\subsection{Simulation Procedure}
The simulation of each exciton is independent and executed as follows.

\begin{enumerate}

\item \textbf{The initial condition:} The 2D wavevector of the exciton at the beginning of each simulation points at a random direction, which we choose according to the uniform distribution $\theta_0 \sim \mathcal{U}[0,2\pi]$.  The angle is measured from the $x$-axis. We have verified that the results we present do not vary measurably when we assume all excitons to have the same angle at $t=0$. The amplitude of the initial wavevector is extracted from the initial kinetic energy, whose value is randomized according to a normal distribution $E_{K,0} \sim \mathcal{N}[E_0,\sigma^2]$. $E_0$ is the average initial kinetic energy  of the exciton and $\sigma^2~=25$~meV$^2$ is the variance.

\item \textbf{Before a scattering event:}  We randomize a value for the free flight duration with the help of the direct technique \cite{Jacoboni:1983jb,Pezzoli_PRB13,Qing_PRB15}
\begin{equation}
\tau_f=-\tau_m \ln(r)\,\,, \label{eq:tau_f}
\end{equation}
where $\tau_m$ was defined in Eq.~(\ref{eq:taum}) and $r \sim \mathcal{U}[0,1]$ is a random number distributed uniformly between 0 and 1.  Eq.~(\ref{eq:tau_f}) guarantees  that the probability for an exciton to scatter in the time interval, $[t,t+dt]$, where $t$ is the elapsed time since the last scattering event, follows an exponential distribution, $X \sim \text{Exp}(1/\tau_m)$. This describes a Poisson process in which scattering events occur continuously and independently at a constant average rate. 

\vspace{2mm}

Next, we check what type of scattering took place after the time of flight, $\tau_f$. To do so we randomize a second number $x \sim \mathcal{U}[0,1]$ distributed uniformly between 0 and 1, and check its value according to the following:
\begin{enumerate}
\item If $0 \leq x < x_1$ where $x_1 = \tau_m/\tau_{E_2', K,+}$, then we elect the emission of LO phonon through the Fr\"{o}hlich scattering.
\item If $x_1 \leq x < x_2$ where $x_2 - x_1 = \tau_m/\tau_{A_1', K,+}$, then we elect emission of ZO phonon through the  short-range interaction with thickness fluctuations.
\item If $x_2 \leq x < x_3$ where $x_3 - x_2 = \tau_m/\tau_{\text{ac}, K,+}$, then we elect the emission of acoustic phonon through the interaction with the deformation potential.
\item If $x_3 \leq x < x_4$ where $x_4 - x_3 = \tau_m/\tau_{\text{ac}, K,-}$, then we elect the absorption of acoustic phonon through the interaction with the deformation potential.
\item If $x_4 \leq x < x_5$ where $x_5 - x_4 = \tau_m/\tau_{\text{r}, K}$, then the exciton recombined radiatively. 
\item If $x_5 \leq x < x_6$ where $x_6 - x_5 = \tau_m/\tau_{\text{nr}, K}$, then the exciton recombined non-radiatively.
\item If $x_6 \leq x < 1$     then we say that the exciton experienced a self-scattering event. 
\end{enumerate}

\item \textbf{After a scattering event}: If the chosen event in step 2 was self scattering then nothing is changed before and after scattering, and we repeat step 2 with the same wavevector. Otherwise, 
\begin{enumerate}

\item We first write into the output file the overall time of the scattering event as well as the pre-scattering amplitude and angle of the exciton's wavevector.

\item The simulation is terminated if the exciton recombined in step 2 or if we already recorded 500 phonon scattering events for this exciton (to save time and space). 

\item If the chosen event in step 2 was scattering with a phonon,  then we repeat step 2 with a new post-scattering wavevector, $\mathbf{K}_{\text{new}} = \mathbf{K}-\mathbf{q}$. In order to do so, we need to choose the phonon wavevector $\mathbf{q}$ that was involved in the scattering. We choose its amplitude by extracting the probability distribution function from Eq.~(\ref{eq:FGR}) after integrating out the angular dependence in the argument of the $\delta$-function
\begin{equation}
f_{\lambda}(q) =   \frac{|M_{\lambda,\mathbf{q}}|^2}{ \sqrt{ 1 - \left( \frac{q}{2K} \pm \frac{K_{\lambda,q}^2}{2Kq}  \right)^2 }}\,\,.  \label{eq:pdf}
\end{equation}
The range of permissible wavevector values, $q_{\text{min}} < q < q_{\text{max}}$, and $K_{\lambda,q}$ are defined as follows.
\begin{enumerate}

\item  If the scattering was through emission of ZO or LO phonons ($\lambda= \{E_2',A_1'\}$), then Eq.~(\ref{eq:pdf}) is taken with the plus sign and
\begin{equation}
q_{\text{min}}  = K - \sqrt{K^2 - K_{\lambda,q}^2 }\,\,,\qquad  q_{\text{max}}  = K + \sqrt{K^2 - K_{\lambda,q}^2}\,\,,\,\,\,\,\,\,\, \text{where}\,\,\,\,\,   K_{\lambda,q} = \sqrt{\frac{2(m_e+m_h)E_{\lambda}}{\hbar^2}}.
\end{equation}

\item  If the scattering was through emission of an acoustic phonon, then Eq.~(\ref{eq:pdf}) is taken with the plus sign and
\begin{equation}
q_{\text{min}}  = 0 \,,\quad  q_{\text{max}}  = \text{max}\left\{0,2\left(K - q_s\right)\right\} \,,\quad  K_{\lambda,q} =\sqrt{2q_sq}  \,,\,\,\,\,\,\,\, \text{where}\,\,\,\, q_s =  \frac{(m_e+m_h)v_s}{\hbar}\,.
\end{equation}

\item  If the scattering was through absorption of an acoustic phonon, then Eq.~(\ref{eq:pdf}) is taken with the minus sign and
\begin{equation}
q_{\text{min}}  = \text{max}\{0,2(q_s-K)\} \,,\quad  q_{\text{max}}  = 2(K + q_s) \,,\quad  K_{\lambda,q} =\sqrt{2q_sq}  \,,\,\,\,\,\,\, \text{where}\,\,\,\, q_s =  \frac{(m_e+m_h)v_s}{\hbar}\,.
\end{equation}
\end{enumerate}

\vspace{2mm}

Next, we use Eq.~(\ref{eq:pdf}) and randomize the amplitude of $q$ by applying the direct technique \cite{Jacoboni:1983jb}.  Specifically, we randomize a number $y \sim \mathcal{U}[0,1]$, and choose the value of $q$ by requiring that
\begin{equation}
y =  \frac{\int_{q_{\text{min}}}^{q} dq' f_{\lambda}(q')}{ \int_{q_{\text{min}}}^{q_{\text{max}}} dq' f_{\lambda}(q') }\,\,.
\end{equation}
Once the value of $q$ is chosen,  we can finally select the amplitude and then the angle of the new wavevector, 
\begin{equation}
K_{\text{new}} =\sqrt{  K^2 \mp K_{\lambda,q}^2 }\,\,\,,\qquad \theta_{\mathbf{K}_{\text{new}}} = \theta_{\mathbf{K}} + \text{sign}(z) \cdot \arccos\left( \frac{ K^2 + K_{\text{new}}^2 - q^2}{2KK_{\text{new}}}\right).
\end{equation}
The $-$ ($+$) sign in the first expression corresponds to the case that the scattering event involved phonon emission (absorption). $z \sim \mathcal{U}[-1,1]$ is a uniformly distributed random number between -1 and 1, whose sign dictates whether we add or subtract the angle change. Having $K_{\text{new}}$ and $\theta_{\mathbf{K}_{\text{new}}}$ we repeat step 2. 
\end{enumerate}

\end{enumerate}

\subsection{Calculation of the depolarization} \label{sec:figs}

\vspace{2mm}

We use a 4-component vector $\mathbf{P}$ to denote the probability of each exciton to belong to one of the four exciton branches and we update this vector after each scattering: $\mathbf{P}_m$ is the probability vector after the $m^{\text{th}}$ scattering event. We assume the initial helicity of the exciton to be $\sigma^+$ (i.e., $| \ell=1 \rangle$ at $t=0$): $\mathbf{P}_0=[1,0,0,0]^T$.  To calculate how $\mathbf{P}$ evolves, we define the transition matrix $\mathbf{Q}$:
\begin{equation} \label{eq:evolve}
\mathbf{P}_m=\mathbf{Q}_m\cdot \mathbf{P}_{m-1}=\prod_{i=1}^{m} \mathbf{Q}_i\cdot \mathbf{P}_0,
\end{equation}
where the $(\ell,j)$ element of the $\mathbf{Q}$-matrix is defined by the probability that the exciton superposition state evolves from branch $| \ell \rangle$ at time $t$ to branch $| j \rangle$ at time $t+\tau_{f,m}$
\begin{equation} \label{eq:Q}
\mathbf{Q}_{m,\ell j}=\left| \langle \ell_t | j_{t+\tau_{f,m}} \rangle \right|^2 =   \left| \sum_n C_{\ell,n}^{\ast}(\mathbf{K}_{m-1}) C_{j,n}(\mathbf{K}_{m-1}) e^{-iE_n(\mathbf{K}_{m-1})\tau_{f,m}/\hbar} \right|^2.
\end{equation}
$\tau_{f,m}$ is the flight time duration between the $(m-1)^{\text{th}}$ and $m^{\text{th}}$ scattering events (excluding the trivial  self-scattering events).  $C_{\ell,n} (\mathbf{K}_{m-1}) \equiv \langle  n, \mathbf{K}_{m-1}   | \ell \rangle=(| n, \mathbf{K}_{m-1} \rangle_\ell)^*$, where $(| n, \mathbf{K}_{m-1} \rangle_\ell)^*$ is the conjugate of the $\ell^{\text{th}}$ element of the $n^\text{th}$ eigenvector of the Hamiltonian $\mathcal{H}(\mathbf{K}_{m-1})$. The  eigenvalues of the latter are $E_n(\mathbf{K}_{m-1})$. 

\vspace{2mm}
The circular polarization degree of an exciton after the $m^{\text{th}}$ scattering event is found from  $\mathbf{P}_m(1)-\mathbf{P}_m(2)$. Finally,  the circular polarization degree of the entire system at time $t$ is calculated by selecting the excitons that radiate during this time.  To that end, we first generate the probability density function (pdf) according to 
\begin{equation} \label{eq:pdf_1}
f(t)=  \frac{1}{T_{N}} \sum_j^N \exp{\left(- \frac{1}{2}  \left(   \frac{t-t_j}{\sigma_N}  \right)^2  \right )} \,,
\end{equation}
where $T_N$ is a normalization factor such that $\int_0^\infty f(t)dt = 1$. $N$ is the number of excitons that end their life through radiative recombination in our simulations. $t_j$ is the time at which the $j^{\text{th}}$ exciton recombined radiatively. $\sigma_N$ is a broadening chosen to bridge between the discrete recombination events and the continuous pdf. Ideally, when $N \rightarrow \infty$, we can choose $\sigma_N \rightarrow 0$ so that the Gaussian becomes a delta function. Practically, however, we have found that the pdf is smooth enough when $N>10^4$ and $\sigma_N$=0.03~ps. The latter is chosen small enough compared with the exciton lifetime.  

Using this method,  the average circular polarization that one can measure in a DC-type experiment follows from 
\begin{equation} \label{eq:p_avg}
\langle p \rangle = \frac{1}{T_N}  \int_0^{\infty} dt \left[ \sum_j^N p_j \exp\left (- \frac{1}{2}  \left(   \frac{t-t_j}{\sigma_N}  \right)^2  \right ) \right] \,, 
\end{equation}
where $p_j$ is the circular polarization of the $j^{\text{th}}$ exciton at the time of its radiative recombination. Equivalently, we can write  $\langle p \rangle =  \int_0^\infty dt p(t) f(t)$, where $p(t)$ is the circular polarization degree of excitons that radiate at time $t$, 
\begin{equation} \label{eq:pt}
p(t)=   \left[ \sum_j^N p_j \exp\left (- \frac{1}{2}  \left(   \frac{t-t_j}{\sigma_N}  \right)^2  \right ) \right] \Bigg/  \left[ \sum_j^N  \exp\left (- \frac{1}{2}  \left(   \frac{t-t_j}{\sigma_N}  \right)^2  \right ) \right] \,.
\end{equation}
Unlike $\langle p \rangle$ and $f(t)$, it is emphasized that  $p(t)$ is unaffected by the exciton lifetime. The reason is that $p(t)$ depends only on the energy relaxation process up to time $t$, while being indifferent to the number of excitons that radiated until time $t$.

Figure~\ref{fig:pANDf} shows the behavior of $p(t)$ and $f(t)$ for the case of ML-MoSe$_2$ ($\Delta_{\text{bd}}=+1.5$~meV). The initial kinetic energy of the exciton is $E_0 = 100$~meV and the temperature is 5~K. We have collected more than 40,000 excitons that ended their life radiatively for the calculations of the time-resolved PL profiles. For the calculation of the polarization decay, $p(t)$, we have also assigned $J_0 = 10$~meV$\cdot \AA$~ for the exchange parameter, while the amplitude of the Rashba parameter was randomized uniformly in the range $0 < \alpha_R E_z < 0.05$~eV$\cdot$\AA$\,$ with each scattering. Given the scarcity of radiative events at early times, we have also ran dedicated simulations to collect radiative events only if they took place at very early times (i.e., when the PL intensity just starts to increase). We have collected these simulations until we were able to generate accurate enough curves for $p(t)$ when $t > 0.1$~ps. In addition, we have assigned $E_{\ell}=3$~meV in Eq.~(\ref{eq:tau_rnr}) for the case that the exciton lifetime is 10~ps, while assigning $E_{\ell}=100$~meV for all other cases. A small $E_{\ell}$ further suppresses the non-radiative process by making it active only for low energy excitons. We see that the polarization decay in Fig.~\ref{fig:pANDf}(a) remains indifferent to these changes. On the other hand, Fig.~\ref{fig:pANDf}(b) shows that the PL profiles are strongly affected by the exciton lifetime.

Next, we calculate the average polarizations and quantum yields in Fig.~\ref{fig:pANDf}. The former is calculated from  $\langle p \rangle =  \int_0^\infty dt p(t) f(t)$, whereas the quantum yield is the ratio between the number of simulations in which excitons ended their life radiatively and the total number of simulations.  Due to the minuscule size of the light cone, radiative events are rare compared with excitons that end their life non-radiatively. When the exciton lifetimes are $\tau_{\text{nr},0}=0.5$, 1.5, 2.5, 5, and 10 ps, we got that $\langle p \rangle \sim22$\%, 12\%, 7\%, 3\%, and 0.7\% , whereas the quantum yields are 0.0054\%, 0.044\%, 0.14\%, 0.74\%, and 5.98\%, respectively. As expected, shorter exciton lifetimes increase the average polarization, but lower the quantum yield because most excitons vanish non-radiatively. Conversely,  longer exciton lifetimes increase the quantum yield but lower  the average polarization.  Finally, we mention that exciton-exciton scattering events should improve the quantum yield by scattering excitons directly to the minuscule  light cone. This effect should be especially relevant when the exciton lifetime is ultra-short (i.e., when there is not enough time to emit multiple phonons to reach the light cone). 

\begin{figure*}
\includegraphics[width=17cm]{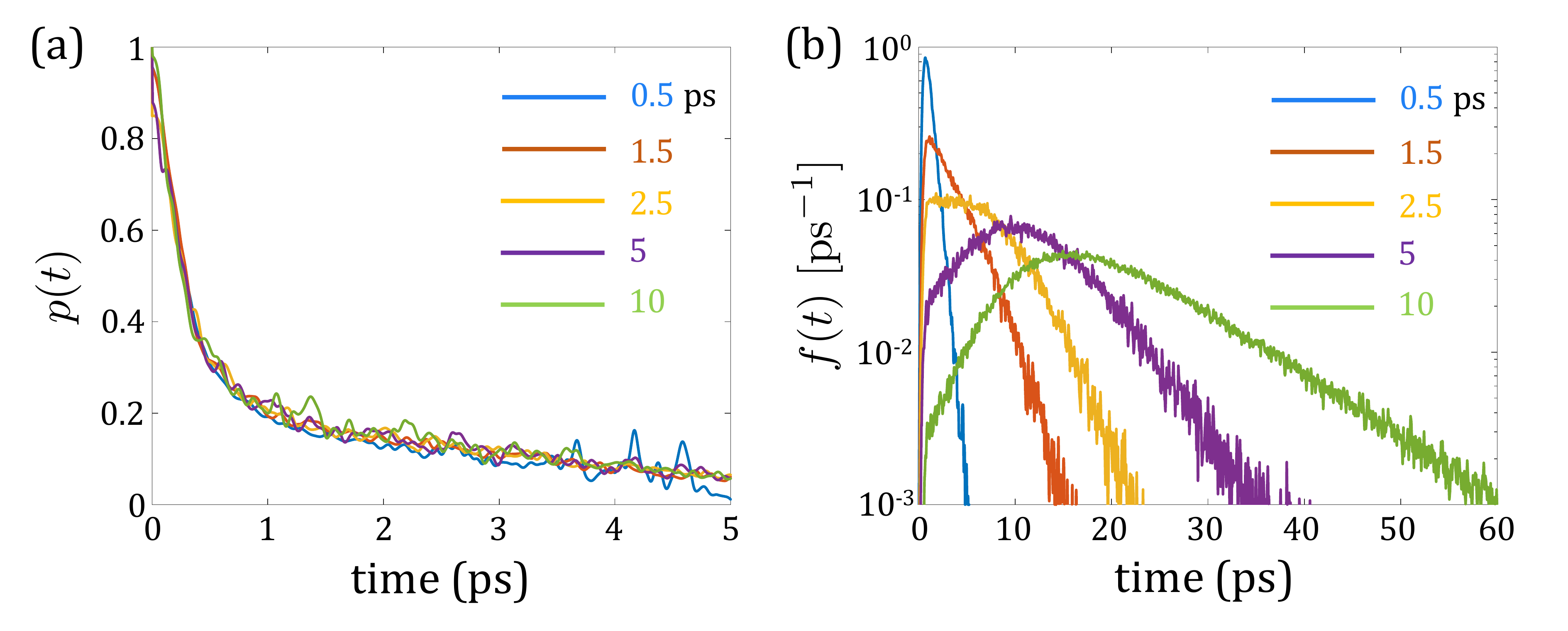}
 \caption{(a) Decay of the circular polarization degree versus time, $p(t)$, for various exciton lifetimes. (b) The probability density function, $f(t)$, extracted from the distribution of radiative events. The shapes of $f(t)$ match well with the PL profiles  measured in ultrafast time-resolved experiments \cite{Robert_PRB16}.\label{fig:pANDf} }
\end{figure*}


\subsection{Figures 2-4 of the main text} \label{sec:figs}
The results in the main paper follow the parameters of ML-MoSe$_2$, while treating $\Delta_{\text{bd}}$ as a free parameter. Using the parameters of ML-WSe$_2$ leads to similar qualitative results. The energy relaxation profile shown in Fig.~2 of the main text relies on the average behavior of 10$^5$ excitons.  The results shown in Figs.~3 and 4 of the main text (polarization decay) rely on simulated excitons that end their life radiatively (in the light cone). We ran simulations until we have collected $4\times10^5$ such excitons.  Actually, we have checked that collecting $\sim 2\times10^4$ radiative excitons is sufficient to observe the salient features. The calculation of $\langle p \rangle$ as a function of the exciton lifetime in the inset of Fig.~3 was calculated by collecting at least $2\times10^4$ for each value of the exciton lifetime. 

In addition, the results shown in Figs.~2-4 of the main text are not affected much by choosing longer radiative times for $\tau_{\text{r},0}$ in Eq.~(\ref{eq:tau_rnr}). The main difference is a lower quantum yield much (and longer computation time). Similarly, choosing a longer exciton lifetime (larger value for $\tau_{\text{nr},0}$) do not affect the results of Fig.~2 and the one shown in the main body of Fig.~3. On the other hand, choosing a longer exciton lifetime (or a few meV for $E_{\ell}$) will result in smaller average polarization values in the inset of Fig.~3 and in Fig.~4(a), while keeping the shape of these curves largely intact. 

\end{widetext}

\end{document}